\documentclass[a4paper,11pt]{article}
\usepackage{pos}

\title{Searching for the X17 with the PADME experiment}

\author{Venelin Kozhuharov$^{a,b}$ for the PADME collaboration \footnote{
The PADME collaboration: 
S. Bertelli, F. Bossi, R. De Sangro, C. Di Giulio, E. Di Meco, D. Domenici, G. Finocchiaro, L.G. Foggetta,
M. Garattini, A. Ghigo, P. Gianotti, M. Mancini, I. Sarra, T. Spadaro, E. Spiriti, C. Taruggi, E. Vilucchi, (INFN
Laboratori Nazionali di Frascati), V. Kozhuharov (Faculty of Physics, University of Sofia “St. Kl. Ohridski”, and
INFN Laboratori Nazionali di Frascati), K. Dimitrova, S. Ivanov, Sv. Ivanov, R. Simeonov (Faculty of Physics,
University of Sofia “St. Kl. Ohridski”), G. Georgiev (Faculty of Physics, University of Sofia “St. Kl. Ohridski”
and INRNE, Bulgarian Academy of Science), F. Ferrarotto, E. Leonardi, P. Valente, A. Variola (INFN Roma1),
E. Long, G.C. Organtini, M. Raggi (Physics Department, ''Sapienza'' Universit`a di Roma and INFN Roma1), A.
Frankenthal (Department of Physics, Princeton University)
}}

\affiliation[a]{Faculty of Physics, Sofia University ``St. Kl. Ohridski'', 
  5 J. Bourchier Blvd., 1164 Sofia, Bulgaria}
\affiliation[b]{INFN - Laboratori Nazionali di Frascati, 
Via Enrico Fermi 54 (già 40) - 00044 Frascati (Roma) Italy }

\emailAdd{venelin@phys.uni-sofia.bg}

\abstract{The PADME experiment was originally designed to test dark matter theories 
predicting the existence of a ''Dark Sector'' 
composed of particles that interact with Standard Model ones 
exclusively through the exchange of a new, massive mediator.
The confirmation of the X17 anomaly, 
observed in nuclear decays at the ATOMKI in Debrecen, 
sparked considerable interest in the particle physics community.
If the anomaly arises from the decay of a new state into an $e^+e^-$ pair, 
the time-reversal symmetry implies that it must be also producible 
through $e^+e^-$  annihilation. 
The PADME experiment can rely on the world's only $e^+$ beam with the 
appropriate energy for a resonant production of X17. 
The collaboration dedicated 2022 data taking 
to investigate the X17 anomaly via $e^+e^- \to X17 \to e^+e^-$  reaction, 
aiming to probe the particle hypothesis.
An overview of the scientific program of the experiment 
and the present status of the search for X17 at PADME are presented.}

\FullConference{42nd International Conference on High Energy Physics (ICHEP2024)\\
18-24 July 2024\\
Prague, Czech Republic\\}

\begin{document}
\maketitle

\section{Introduction}

Despite the great success of the Standard Model (SM)
of particle physics several 
major astrophysics observational phenomena  still miss a
consistent explanation, 
most notably, the origin of the matter-antimatter asymmetry 
in the Universe and the nature of Dark matter. 
In addition, 
there are  a few smoking guns among the particle physics 
experimental results 
still in tension with the present understanding. 
This includes the long time existing discrepancy in the 
anomalous magnetic moment $a_{\mu}$ of the muon \cite{Muong-2:2023cdq}
and the observation of a structure in the 
angular distribution of $e^+e^-$, 
emitted through internal pair creation (IPC) in 
excited $^8$Be, $^4$He and $^{12}$C nuclei \cite{Krasznahorkay:2015iga, Krasznahorkay:2021joi, Krasznahorkay:2022pxs}. 

All mentioned observations may 
have their solution in various extensions of SM, 
which predict the existence a new hidden sector of particles 
accessible only through a portal. 
In the simplest case the portal could be due to a new U(1) gauge symmetry, 
with a vector mediator \cite{holdom2}. 
Then the interaction of the new vector particle with the 
SM fermions is generally governed by a $g_{vf}$ coupling, 
which may also arise effectively through mixing with the ordinary photon. 
In the case of IPC in $^8$Be, $^4$He and $^{12}$C, 
the preferred mass of the hypothetical intermediate state particle, the X17, 
is $M_{X17} = 16.85 \mathrm{MeV}$ \cite{Denton:2023gat}, 
a value which  
could also be consistent with the measured $a_{\mu}$ \cite{Muong-2:2023cdq}.

\section{The PADME experiment at LNF-INFN}

\begin{figure}
\centering
\includegraphics[width=\textwidth]{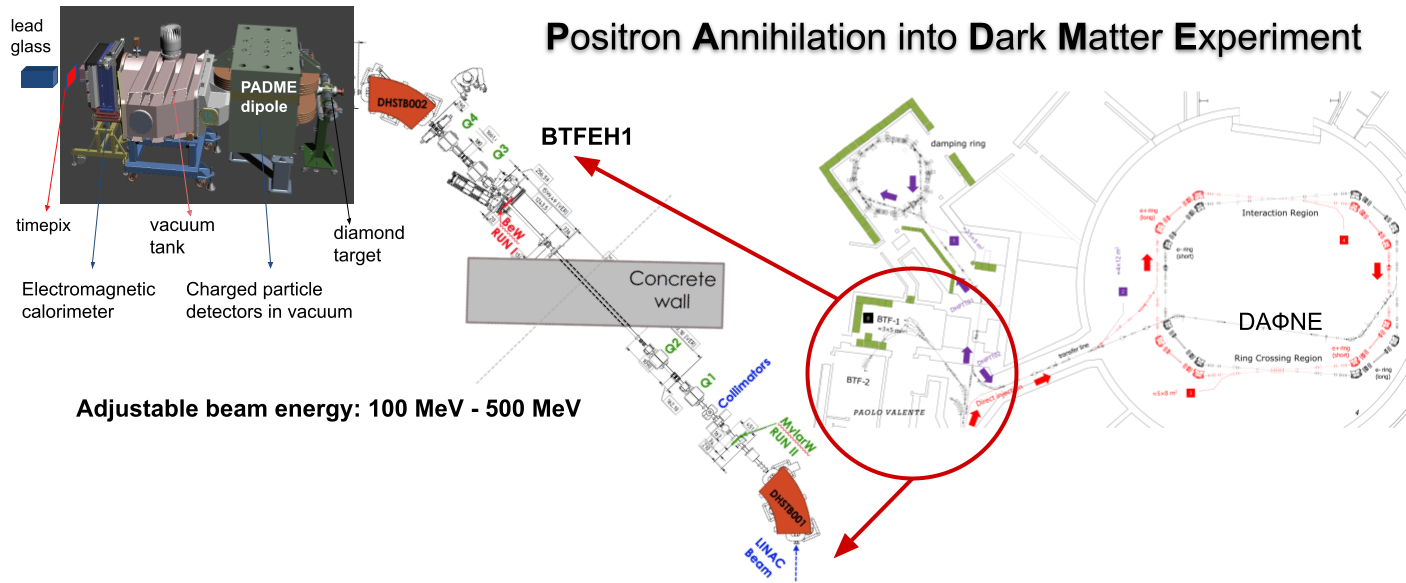}
\caption{A sketch of the PADME experiment together with the DA$\Phi$NE accelerator complex. 
The DHSTB001 dipole magnet deflects the positrons to BTFEH1 while DHSTB002 directs them 
towards the PADME target.  
\label{fig:padme-setup}}
\end{figure}

The Positron Annihilation into Dark Matter Experiment (PADME) \cite{Raggi:2014zpa}
aims
to search for new light states 
with mass below O(20) MeV in 
positron-on-target annihilation process. 
PADME is located at Laboratori Nazionali di Frascati and
utilizes positron beam with energy 
up to 490 MeV
$E_{beam}$ 
from the DA$\Phi$NE
Linear accelerator, as shown in fig. \ref{fig:padme-setup}. 
The accelerated positrons are deflected by a 
pulsed dipole magnet 
to a dedicated experimental hall, BTFEH1, 
where they are directed towards the PADME 
setup by an additional dipole magnet. 
The positron energy can be varied from 
O(100) MeV to about 500 MeV, when positrons are produced 
by a secondary target at the exit of the accelerator chain, 
or to about 430 MeV when a positron converter located after 
the first electron accelerator stage is used. 
The Linac provides 50 bunches of particles per second, 
one of which is deflected towards a spectrometer for energy monitoring
and 49 are delivered to users.

The positron beam impinges onto a 
100 $\mathrm{\mu m}$ thick diamond target
with size 2 cm $\times$ 2 cm \cite{Chiodini:2017uzx}. 
The ionisation charge in the diamond is readout
by engraved by excimer laser horizontal and vertical strips. 
An evacuated region placed inside 
a 0.5 T dipole magnet follows. 
The vacuum tank hosts three sets of 
charged particle detectors, positron veto (PVeto), electron veto (EVeto) 
and the high energy positron veto (HEPVeto) 
made of plastic scintillator bars and 
readout by Hamamatsu S13360 silicon photomultipliers,
which allow the detection 
of lower energy positrons 
due to hard bremsstrahlung emission in the target \cite{PADME:2024suz}.
The neutral products of the interaction of the positrons in the 
target are detected by a ring-shaped segmented electromagnetic calorimeter 
(ECal, \cite{Albicocco:2020vcy})
made of 616 BGO crystals, read out by HZC 1911 photomultipliers. 
The inner hole of the ECal is covered by a Cherenkov calorimeter 
composed by $5\times5$ matrix of 
PbF$_2$ crystals, the SAC (Small Angle Calorimeter) \cite{Frankenthal:2018yvf}, 
which allows to disentangle the high rate of bremsstrahlung photons
with $E_{\gamma} \geq 50 \mathrm{MeV}$. 

PADME started taking data in the autumn, 2018 
with secondary positron beam with 
multiplicity of 20000 positrons per bunch and 
energy 
$E_{beam} \simeq 490~\mathrm{MeV}$ for
an $e^+e^-$ invariant mass $M_{e^+e^-} = 22.4 ~\mathrm{MeV}$. 
The recorded data, so-called RUN I, was extensively used for
detector calibration and commissioning of the experimental setup 
and 
for understanding and developing methodology for background suppression. 

In the autumn, 2020 a three month long data taking period 
was initiated, the so-called RUN II, 
with primary positron beam with energy
$E_{beam} \simeq 430~\mathrm{MeV}$
and an upstream Mylar window instead of a beryllium one 
to protect the Linac vacuum. 
Although with lower beam energy leading to a reduced 
invariant mass ($M_{e^+e^-} = 21 ~\mathrm{MeV}$) 
and thus slightly reduced access to 
the new light particles parameter space, 
the primary positron beam 
exhibited much smaller
beam induced background. 
Both RUN I and RUN II were devoted to the search for associate
production of new states $A'$ in the process $e^+ + e^- \to A' + \gamma$. 
Upon the detection of the recoil 
photon, the missing mass squared, $M_{miss}^2 = (P_{e^+} + P_{e^-} - P_{\gamma})^2$ 
is used as a signal discriminator,
where $P_{e^+}$, $P_{e^-} = (m_e, 0,0,0)$, $P_{\gamma}$ 
are the positron, electron and the measured gamma four-momenta, 
respectively.

RUN III took place in autumn, 
2022 with a modified experimental setup, 
dedicated to the search for X17 in
the so called
resonant production mode \cite{Darme:2022zfw}. 
Due to the expected enhanced X17 production cross-section 
at resonant beam energy, 
the beam intensity was limited to  
$\leq$ 3000 positrons per bunch, 
decreasing significantly the 
beam induced background. 
The dipole magnet together with EVeto, PVeto and HEPVeto
were switched off, 
a new charged particle detector (ETagger)
made of plastic scintillators and readout by 
SiPMs was placed in front of the ECal, 
SAC was removed and
a matrix of $2 \times 6$ 
Timepix3 silicon pixel detector array \cite{Bertelli:2024jzb}
was 
placed downstream of the ECal to measure the beam 
size and position at the exit of the experimental setup. 
In addition, a lead glass block placed 
after the Timepix3 served as a beam dump and 
measured the beam multiplicity by measuring the total deposited energy
for each bunch \cite{PADME:2024pwg}.

\section{Searching for new light particles}

PADME executed two separate search campaigns 
for new light particles exploiting the missing mass technique 
(RUN I and RUN II) by reconstructing single photon final states, 
and the resonant X17 production technique. 
While the analysis of the single photon events in RUN I and RUN II
is still in progress and profits from the development 
of new methods for the mitigation of higher ECal occupancy \cite{Dimitrova:2022uum}, 
the analysis of a fraction of the PADME data of events with 
two clusters in the ECal lead to the most precise measurement 
of the semi-inclusive cross-section 
for the process $e^+e^- \to \gamma \gamma$ \cite{PADME:2022tqr} 
at $E_{beam} = 430 ~\mathrm{MeV}$:
\begin{equation}
    \sigma(e^+ + e^- \to \gamma \gamma) = (1.977 \pm 0.018_{stat} \pm 0.119_{syst}) ~\mathrm{mb}. 
\end{equation}

To probe the X17 existence,
PADME performed a scan on the $e^+ e^-$ invariant mass
profiting from  the significant increase of the 
X17 production
cross-section when $M_{e^+ e^-} \sim M_{X_{17}}$ \cite{Arias-Aragon:2024qji}
and
the possibility to vary the 
DA$\Phi$NE Linac beam energy.
Thus the contribution of X17 is seen as a change in the 
$\sigma(e^++e^- \to e^+ + e^-)$ cross-section as a function of $\sqrt{s}$. 

\begin{figure}
\centering
\includegraphics[width=0.49\textwidth]{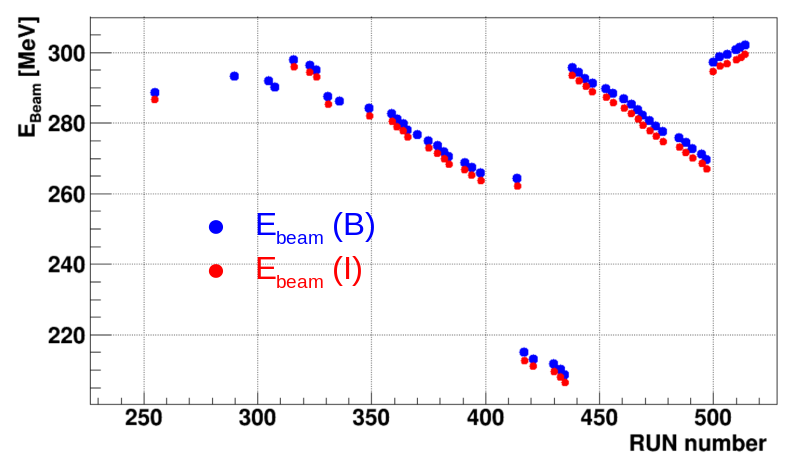}
\includegraphics[width=0.49\textwidth]{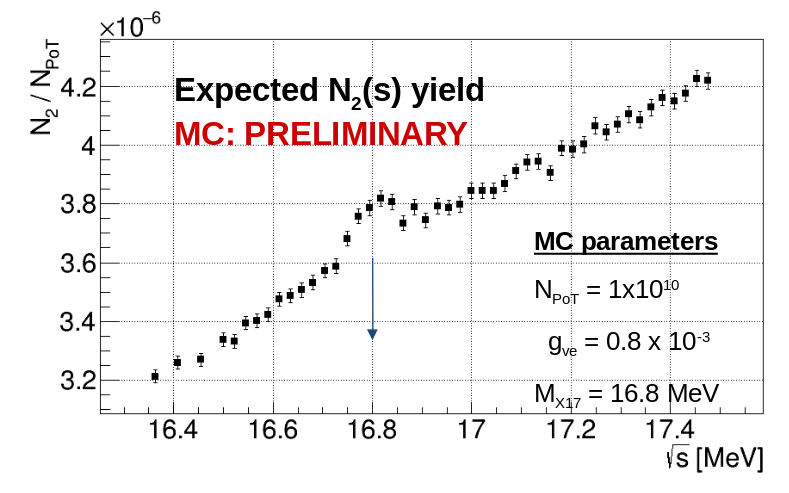}
\caption{Left: A summary of the PADME runs taken at different beam energy, 
measured by monitoring the dipole magnet current and by Hall probes. Right: Expected rate of 
the reconstructed two cluster events in the case of a 
presence of a new state with mass $M_{X_{17}} = 16.8~\mathrm{MeV}$ with 
coupling $g_{ve} = 0.8\times 10^{-3}$.
\label{fig:padme-energy-events}}
\end{figure}

An energy scan was performed, collecting data at 47 points with 
positron beam energy
$263~ \mathrm{MeV} \leq E_{beam}\leq 299~\mathrm{MeV} $, 
spaced by $0.7~ \mathrm{MeV}$. This provided access to  
$16.4~\mathrm{MeV} \leq M_{X_{17}} \leq 17.5~ \mathrm{MeV}$.
In addition, several data samples with lower ($\sim$ 210 MeV) and 
higher (402 MeV) beam energy 
were recorded for calibration and systematic assessment purposes.
The beam energy was monitored through the current of the DHSTB001 dipole
and by a hall probe. The data samples (runs) with different positron 
beam energy are illustrated in fig. \ref{fig:padme-energy-events}, left. 
For each energy point about $10^{10}$ positrons on target were collected. 

With the PADME dipole magnet off during the entire RUN III, 
both $e^++e^- \to e^++e^- $ and $e^++e^- \to \gamma \gamma$ final states were 
detected as two cluster events by the ECal. 
Since $\sigma(e^++e^- \to e^+ + e^-) \gg \sigma(e^++e^- \to \gamma \gamma) $,
the change in the total number of 
all two cluster events $N_2$ with the $\sqrt{s}$
was used to probe the existence of X17.
For example, the number of the expected two cluster 
events as function of $\sqrt{s}$ for an X17 mass 
of 16.8 MeV and X17 coupling to electrons $g_{ve} = 0.8 \times 10^{-3}$
is shown in fig. \ref{fig:padme-energy-events}, right.
The analysis strategy is to discriminate
\begin{equation}
    N_2(s) = N_{PoT}(s)\times [B(s) + S(s; M_{X_{17}},g_{ve}) \varepsilon_S(s)]~~ \mathrm{versus} ~~ N_2(s) = N_{PoT}(s)\times B(s), 
\end{equation}
where  $N_2(s)$ ($s=\sqrt{s}^2$) is the number of two cluster events,
$B(s)$ is the background yield per positron on target, 
$N_{PoT}(s)$ is number of the positrons on target,
$\varepsilon_S(s)$ is the signal selection efficiency,
$S(s; M_{X_{17}},g_{ve})$ is the signal production strength as function of the X17 mass, 
coupling and $e^+e^-$ invariant mass,
and $\sqrt{s}$ determined run by run from the  measured
beam energy.
To decrease bias, $N_2(s)$  is kept blind throughout the analysis. 
All geometry cuts were performed with respect to the 
positron beam position at the ECal plane, 
determined by the center of gravity 
of two cluster events and validated by the Timepix3 data.
The beam multiplicity for each energy point was measured by
the lead glass block, calibrated extensively and taking into account the 
transversal energy leakage due to the change of the beam position 
and beam size at the Timepix3 run by run.

\begin{figure}
\centering
\includegraphics[width=0.55\textwidth]{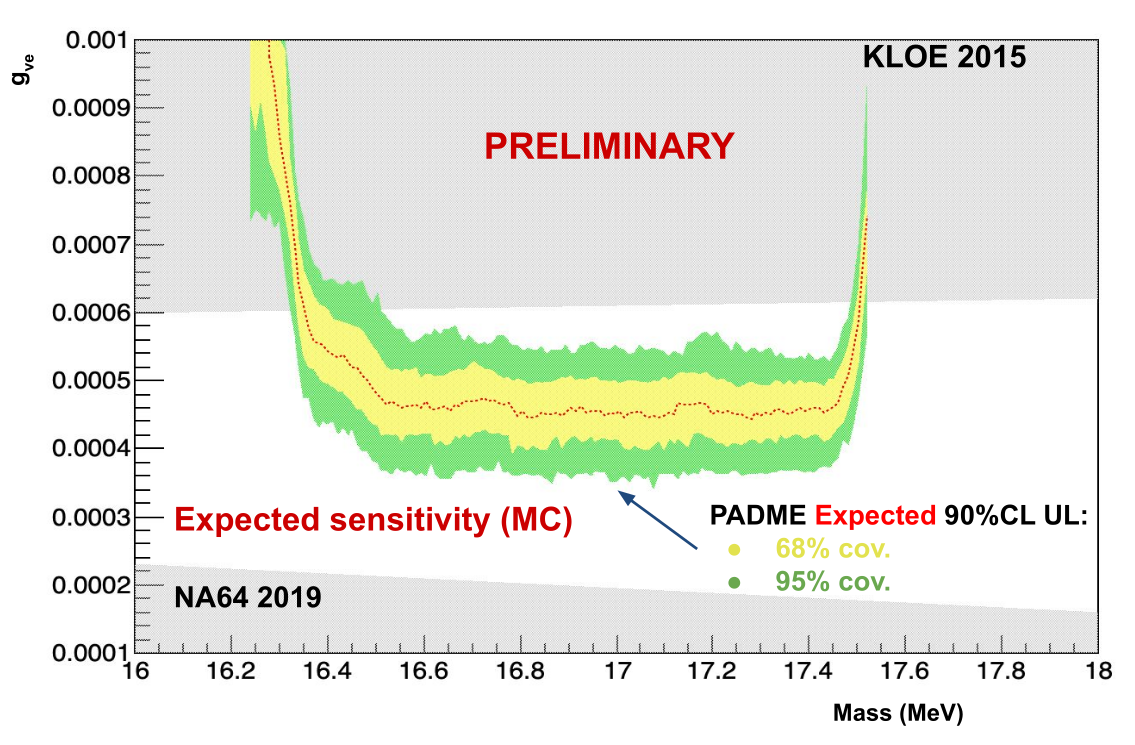}
\includegraphics[width=0.4\textwidth]{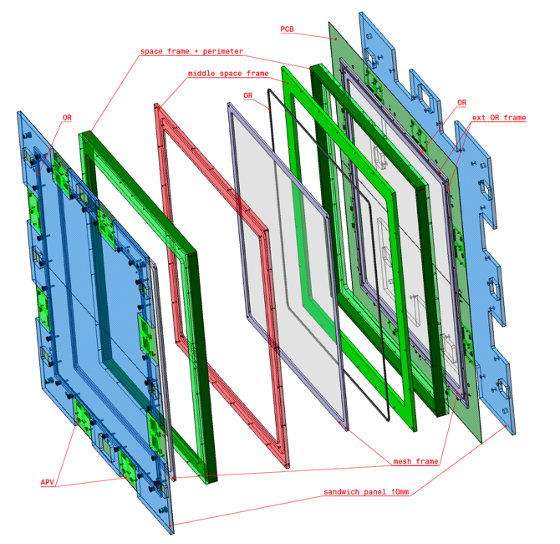}

\caption{Left:Preliminary estimation of the PADME sensitivity (90\% CL limits) to 
the X17 parameter space with RUN III data. Excluded regions 
by KLOE \cite{Anastasi:2015qla} and NA64 \cite{NA64:2019auh} are also presented. 
Right: A sketch of the new PADME Micromegas chamber
to discriminate between charged and neutral 2 cluster final states and to provide measurement 
of the charged particle directions.
\label{fig:padme-X17-prel-sensitivity}}
\end{figure}

The expected PADME sensitivity with the described analysis technique applied 
to the RUN III data is shown in fig. \ref{fig:padme-X17-prel-sensitivity}, left. 
As can be seen, 
PADME will probe completely unexplored region 
with the already collected data. 

The dominant limitations in the analysis of PADME RUN III data 
were identified to be the precise knowledge of the number of positrons on target 
and the two cluster events acceptance systematics. 
Both can be addressed by selecting a different normalization channel, 
$e^+ e^- \to \gamma \gamma$, and measure the relative cross section 
$\sigma(e^+ e^- \to e^+ e^-)/\sigma(e^+ e^- \to \gamma \gamma)$ as a function of
$\sqrt(s)$. 
The similarity in the topology cancels to high extent the 
acceptance related effects and the relative ratio does not suffer
from absolute beam flux determination.
For precise charged particle identification
and
discrimination between $e^+ e^- $ and $\gamma \gamma$ final 
states, a new Micromegas based detector (fig. \ref{fig:padme-X17-prel-sensitivity}, right)
with a central anode plane and two signal readout planes
is foreseen to be placed in front of the ECal. 
The necessary 0.5 \%  uncertainty in the 
$e^+ e^- \to \gamma \gamma$ sample
will be achieved by collecting larger number of 
positrons-on-target per energy scan point. 
Preliminary estimations indicate that with 
the proposed upgrade and six months of data 
taking PADME will be able to probe the 
entire allowed X17 parameter space. 
A run with the described improvements 
was scheduled and will take place in 2025.

\section{Conclusions}
The PADME experiment at LNF-INFN 
is probing the existence of new states with mass below 
$\sim$22 MeV in $e^+$ on target annihilation process, 
collecting more than $10^{13}$ positrons on target
in three data taking campaigns. 
While RUN I and RUN II data exhibits 
higher occupancy in the electromagnetic calorimeter, 
the analysis of the data dedicated to the search 
for resonant production of X17 is at advanced stage 
and the expected sensitivity covers so far 
unexplored X17 parameter space. 
PADME foresees a new run in 2025 with an upgraded
setup which will allow to completely probe the 
allowed parameter space for $16.4~\mathrm{MeV}\leq M_{X_{17}} \leq 17.4~\mathrm{MeV}$.


\begin{acknowledgments}
The authors acknowledge major support from Istituto Nazionale di Fisica Nucleare, Italy.
Sofia University team acknowledge that partially this study is financed by the
European Union-NextGenerationEU,
through the National Recovery and Resilience Plan of the Republic of Bulgaria,
project SUMMIT BG-RRP-2.004-0008-C01 and by 
BG-NSF KP-06-DO02/4 from 15.12.2020 as part of MUCCA, CHIST-ERA-19-XAI-009.
\end{acknowledgments}

\end{document}